# Perfect Plane-Wave to Surface-Wave Coupler Enabled Teleporting Conformal Metasurfaces

Jordan Budhu, *Member, IEEE*

*Abstract*—A technique for the design of conformal metasurfaces with two spatially disconnected space wave ports connected by a surface wave is presented. The passive and lossless metasurface absorbs the incident plane wave at port 1, converts it perfectly into a surface wave which transports the energy along an arbitrarily shaped/curved metasurface to port 2, then reradiates the captured power as a radiated field with control over its amplitude and phase. Since the incident field is seen to disappear at the input port and reappear at a spatially dislocated port as a new formed beam, the field can be said to have teleported. The metasurface consists of a single, conformal, spatially variant, impedance sheet supported by a conformal grounded dielectric substrate of the same shape. It is modeled using integral equations. The impedances of the sheet are optimized using the adjoint variable method to achieve the perfect teleporting operation from a passive and lossless metasurface. Possible applications include channel optimization for cellular networks, inexpensive power harvesting, sensing, around-the-corner radar, and cloaking.

*Index Terms*—Conformal, Metasurface, Grating coupler

## I. INTRODUCTION

THE design of metasurfaces to create tunnel-like connections through space (with a finite travel time from port to port) connecting two space wave ports at distant locations is addressed in this paper (see Fig. 1). The incident plane wave field is absorbed at port 1, perfectly converted into a surface wave which connects the two ports and transfers power between them, and reradiated from port 2 located at a distant location. The reradiated field from port 2 is designed with arbitrary control over its phase *and* amplitude in a completely passive and lossless way utilizing all of the power contained in the incident field over port 1. As the metasurface transfers all of the available power in the incident wave to the reradiated wave, the operation is said to be perfect.

The enabling technology for the presented designs is perfect plane-wave to surface-wave couplers. Although these devices have been demonstrated before for planar [1]–[4] and cylindrical surfaces [5]–[7], perfect coupling over an arbitrarily shaped non-canonical conformal surface has not been shown. Furthermore, complex-valued field control over non-canonical conformal surfaces using passive and lossless metasurfaces has also not been shown. In this paper, perfect plane-wave to surface-wave couplers enable teleportation of incident beams to spatially dislocated ports along any desired shape surface and with complex-valued field control of the reradiated beam from a completely passive and lossless metasurface.

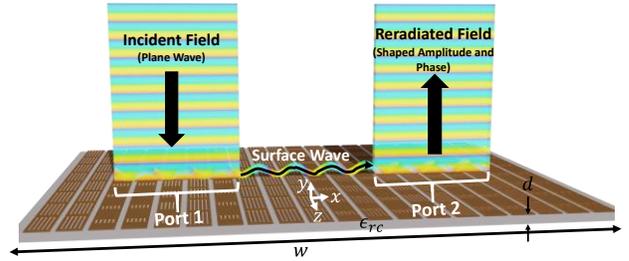

Fig. 1. Teleporting metasurface geometry. Note, the unit cells are shown enlarged for clarity. The actual metasurface contains 200 unit cells each $\lambda_0/20$ wide.

The metasurface itself is a textured interface, modeled as a spatially variant homogenized purely reactive impedance sheet supported by a grounded dielectric substrate [8]. The metasurface is capable of complex-valued radiated field control and seamless conversion between guided and unguided modes in a lossless or perfect manner. It is modeled using integral equations [9]–[11], the integral equations are solved using the method of moments technique [12], and the reactances of the impedance sheet optimized using the adjoint variable method [11], [13], [14]. For an overview of the design procedure, see [11], [15]. A customized integral equation for the design of the conformal cases in this paper can be found in the supplementary material of [16]. The metasurface can be made to conform to any shape such as the corner of a building (see Fig. 7 for preview) or a general non-canonical curvilinear surface (see Fig. 13 for preview) for example. We will first present a planar design to demonstrate the perfect teleportation and understand its operation. Subsequently, the same functionality from conformal geometries will be shown.

Similar planar teleporting metasurfaces have appeared in recent scientific works. In [2], a planar teleporting metasurface was designed by juxtaposing three separate metasurfaces, two space wave to surface wave couplers separated by a metasurface supporting a pure surface wave. The overall metasurface system laterally shifts a plane wave incident at an angle of 30° on the left-hand side of the metasurface to a transmitted wave emanating from the right-hand side of the metasurface at an angle of −7.2° with respect to the normal. Due to the analytical design procedure, the metasurface junctions scatter and reduce the efficiency. Also, the coupling metasurfaces do not perfectly convert the incident fields to surface wave fields. The authors report an efficiency of only

Jordan Budhu is with Virginia Tech, Blacksburg, VA 24060 USA. (e-mail: jbudhu@vt.edu).

10%, and hence the teleportation cannot be deemed perfect. Furthermore, the approach cannot control both the phase and amplitude of the transmitted field and hence does not have the capability of complex-valued field control. The metasurfaces in the referenced work are also not conformal.

In [17], a planar teleporting metasurface is designed using the principles of $\mathcal{PT}$ symmetry. A reactive layer is sandwiched between an absorbing lossy Salisbury screen layer matched to free space and an active layer with a negative impedance also matched to free space. An incident plane wave is absorbed nearly completely by the lossy layer, while the inductive perforated layer allows the remaining small amount of power to couple to the active layer where it is resonantly amplified to recreate or teleport the incident plane wave to the opposite side. Although this device requires active and lossy components, the loss and gain is balanced according to $\mathcal{PT}$ symmetry and hence represents an overall lossless system. Nonetheless, the structure is planar, requires active layers which complicates fabrication and cannot achieve beamforming.

This paper is organized as follows. In section II, we present a planar example. Next, in section III, two conformal examples will be presented. The first contains planar coupling regions and a conformal surface wave region. The second contains both conformal coupling regions and conformal surface wave regions. Some concluding remarks are provided in section IV. An $e^{j\omega t}$ time convention is assumed and suppressed throughout the paper.

## II. Teleporting Metasurface Design and Analysis

We first present a planar teleporting metasurface to understand the teleportation function as it pertains to metasurfaces. The metasurface geometry is shown in Fig. 1. The electromagnetics problem is 2-dimensional (out-of-plane wavenumber is zero) and hence the geometry is invariant in the $z$-direction. The patterned metallic cladding, described by a spatially variant homogenized sheet impedance $\eta_s(x)$, is supported by a grounded dielectric substrate of thickness $d = 1.27\text{mm}$ (50mil) and complex relative permittivity $\epsilon_{rc} = \epsilon_r(1 - jtan\delta) = 2.2 - j0.002$. The impedance sheet is broken into 200 unit cells of width $\lambda_0/20$ each at $f = 10\text{GHz}$. The metasurface is therefore $w = 200(\lambda_0/20) = 10\lambda_0$ wide along the $x$-axis. The design of the teleporting metasurface begins by specifying the desired total field tangential to the metasurface. The incident field is assumed a normally incident plane wave illuminating only the right-hand portion of the metasurface between $\lambda_0 \leq x \leq 4\lambda_0$ as shown in Fig. 2a. The reradiated (scattered) field is defined to have both a cosine tapered amplitude and uniform phase (complex-valued field control) exiting from only the left-hand portion between $-4\lambda_0 \leq x \leq -\lambda_0$. Critically, the absolute level of the amplitude in V/m of the scattered field is chosen to conserve power globally meaning the total power in the incident field, $P_{inc} = |E_0|^2/2\eta_0 = 0.12$ mW/m for a unit strength plane wave, is equal to the total power in the scattered field. This definition will lead to perfect teleportation, i.e., all the power is transferred from the incident field to the scattered field.

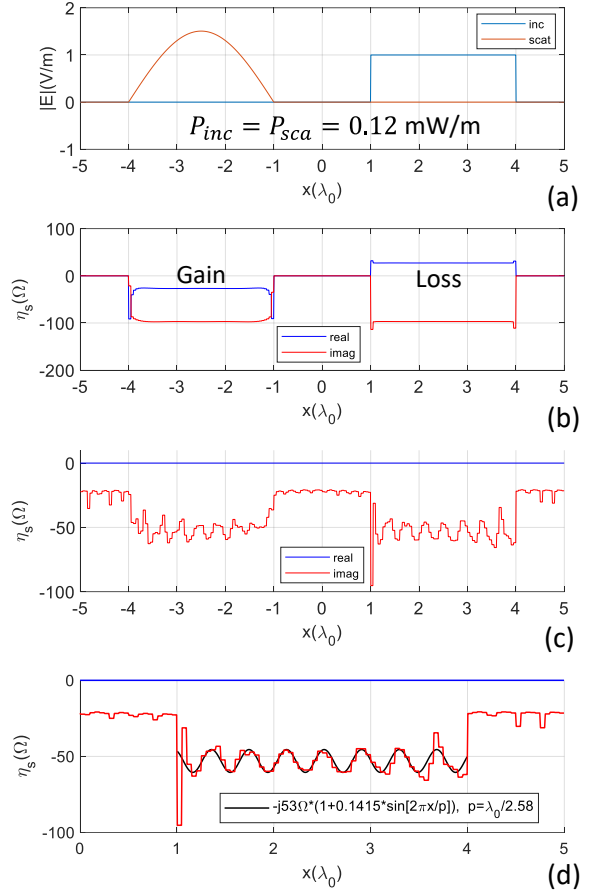

Fig. 2. (a) Specification of incident and scattered field amplitudes at the metasurface plane. NOTE: the input/output ports have been swapped with respect to Fig. 1. Metasurface sheet impedances of (b) Initial local active/lossy metasurface design. (c) Subsequent non-local passive and lossless design. (d) Zoomed in view within the input port region of the non-local passive and lossless design shown superimposed with the sheet impedance modulation function $\eta_s = -j53\Omega\left[1 + 0.1415\sin\left(\frac{2\pi x}{p}\right)\right]$ where $p = \lambda_0/2.58$.

### A. Local Active/Lossy Design

The metasurface design algorithm [11] starts from a local active/lossy design used as a seed for the non-local passive/lossless design. The local active/lossy design is obtained from the solution of the governing integral equations given the desired total field, $E^{tot}$, associated with the wavefront transformation. By solving the integral equation, the surface current density, $J_s$, on the metasurface is obtained, thereby allowing for the direct calculation of the metasurface impedances, $\eta_s$, following from the boundary condition $\eta_s = E^{tot}/J_s$ (see Fig. 2b). As expected, the metasurface is lossy over the incident field region, and contains gain in the scattered field region. Furthermore, the loss and gain is balanced. A transmission line model can be used to understand the result. Modelling the panel as a transmission line terminated in a shunt impedance representing the metasurface in parallel with an

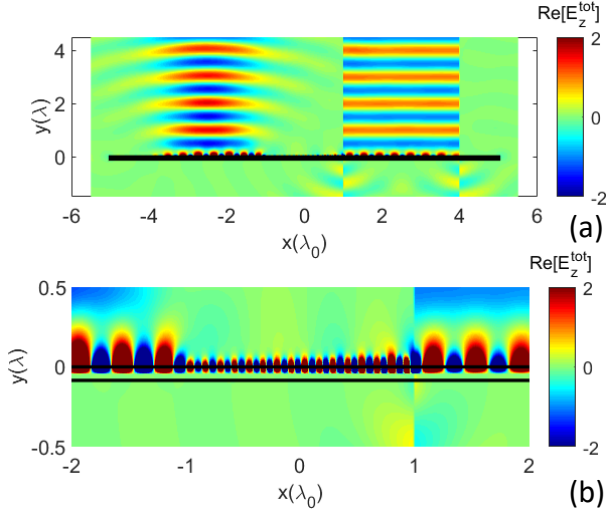

Fig. 3. (a) Real part of the total electric field. (b) Zoomed in to show surface wave connecting the input and output ports.

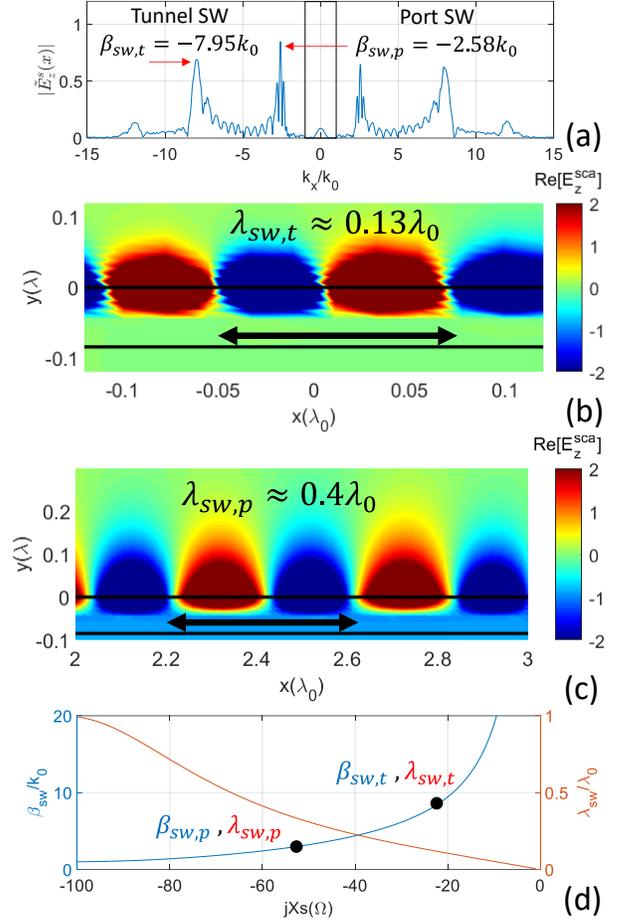

Fig. 4. (a) Amplitude spectrum of the scattered electric field at the metasurface plane. (b) Zoomed in view of Fig. 3a within the tunnel region to show tunnel surface wave wavelength. (c) Zoomed in view of Fig. 3a within the input port region to show port surface wave wavelength. (d) Dispersion curve relating the surface wavenumber and wavelength to the homogenized sheet reactance.

inductance representing the thin grounded dielectric substrate, the sheet impedance can be calculated as

$$\eta_{s,inc} = -\frac{\eta_0 \eta_d \tan \beta d \left(1+\Gamma\right)}{\eta_d \tan \beta d \left(\Gamma - 1\right) - j\eta_0 \left(1+\Gamma\right)}\Bigg|_{\Gamma=0, e^{j0}} = 27.54 - j98\,\Omega \quad (1)$$

where $\eta_{s,inc}$ is the sheet impedance within the illuminated portion of the metasurface, $\Gamma$ is the reflection coefficient looking into the parallel load, $\beta$ is the wavenumber in the dielectric region, and $\eta_0$ and $\eta_d$ are the intrinsic impedances of the free space and dielectric regions, respectively. For perfect absorption, the reflection coefficient should be zero. The resulting sheet impedance in (1) matches the numerically obtained value in Fig. 2b. In order for power conservation, the power absorbed in the lit region must exit the output region, and hence the sheet impedance in the scattered field region can be described as $\eta_{s,sca} = -27.54 - j98\,\Omega$. Note, since the shape of the amplitude differs between the two port regions, the sheet impedance tapers are different at the ends of their respective regions. Thus, the desired functionality of teleportation can be achieved with a local balanced active/lossy metasurface described by the sheet impedances in Fig. 2b. In this case, the beam truly teleports as the incident energy is not transported to the output region but rather the output beam is created through resonant amplification given some small diffractive coupling. Its balanced loss and gain operating principle is similar to the balanced loss and gain of the $\mathcal{PT}$ symmetric teleporting structure in [17]. In both cases, although the incident energy itself does not teleport, a small diffractive coupling is necessary to excite the gain medium to resonantly create the teleported beam.

### B. Non-local Passive/Lossless Design

A purely passive/lossless metasurface eases fabrication and avoids unnecessary complexity. To that end, the local active/lossy metasurface is used as a seed design to obtain a non-local passive/lossless design with the same performance. Since the power is balanced, a surface wave can carry the power from the lossy region to the active region. In this case, the metasurface tunnels the energy to the output port rather than teleports it. However, from an outside perspective, the operation is the same. The integral equation solver is coupled with an adjoint variable optimizer to obtain the non-local design [11]. The non-local metasurface sheet impedance is shown in Fig. 2c and the near fields computed from the non-local design are shown in Fig. 3. Three distinct regions are evident, a spatially modulated input and output port connected by a nearly constant surface wave region. Sharp discontinuities in the sheet impedance of Fig. 2c excite a number of auxiliary surface waves (different from the tunnel surface wave connecting the two ports) in the input and output port regions responsible for distributing power transversally within the port region facilitating passivity and losslessness [11], [18]. These perturbations also aid in obtaining a seamless transition region between the ports and the connecting surface wave region increasing the overall port-to-port power transfer efficiency.

The surface waves can be visualized in Fig. 4a. Fig. 4a shows the amplitude of the plane wave spectrum of the scattered electric field evaluated on the metasurface. The tunnel surface wave responsible for transporting power between the ports is evident at $\beta_{sw,t} = -7.95k_0$. This wavenumber is in agreement with the sheet impedance of $\eta_s = -j22\Omega$ in the tunnel region in Fig. 2c since a sheet of this impedance supports a surface wave of wavenumber $\beta_{sw,t} = -7.95k_0$. This can be verified by viewing the dispersion curves plotted in Fig. 4d. In Fig. 4d, a plot of the surface wavenumber and wavelength versus the sheet reactance of the metasurface is shown. The plot is obtained using the Transverse Resonance Technique [11]. A reactive sheet impedance of $\eta_s = -j22\Omega$ is seen to support a surface wave of wavenumber $\beta_{sw,t} = -7.95k_0$. The surface wave wavelength can also be verified to agree with the curves of Fig. 4d. In Fig. 4b, a zoomed in view of the surface wave in the tunnel region is shown. The measured wavelength agrees with the dispersion curves.

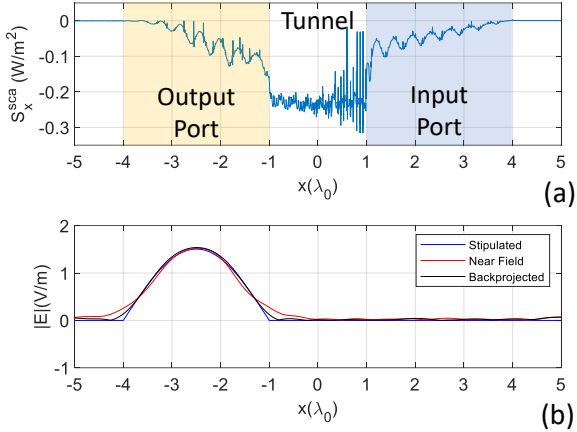

Fig. 5. (a) $x$-component of the Poynting vector at the plane of the metasurface. (b) Line cut of near electric scattered field amplitude at a height of one wavelength above the metasurface.

The amplitude spectrum in Fig. 4a also shows another peak at $\beta_{sw,p} = -2.58k_0$, which is the surface wave generated from the incident plane wave within the input port region. Figure 4d shows this surface wavenumber is associated with a sheet impedance of $\eta_s = -j53\Omega$ in agreement with the average of the sheet impedances shown in Fig. 2c within the input port region. The sheet impedance modulation in the input port region can be understood by noting that for broadside radiation of the $n = -1$ harmonic from a surface wave of wavenumber $\beta_{sw,p} = -2.58k_0$, the period of the modulation should be $k_{xn} = \beta_{sw,p} - 2\pi/p \Rightarrow 0 = \beta_{sw,p} - 2\pi/p$ or $p = 2\pi/\beta_{sw,p} = \lambda_0/2.58$. Shown in Fig. 2d, a sinusoidal sheet impedance modulation function with this period is fit to the non-local passive/lossless metasurface sheet reactances. As can be seen, the modulation period corresponding to the $n = -1$ harmonic for broadside radiation fits the optimized sheet reactances well. The perturbations of the reactances around this analytic result leads to the perfect coupling. No other spatial harmonics fall within the light cone. A zoomed in view of the generated surface wave within the input port region is shown in Fig. 4c. The measured surface wave wavelength is also in agreement with the dispersion curves in Fig. 4d.

Finally, a remark on the spectrum limits. The spectrum has a cut-off at $\beta_{sw} = 10k_0$ which is the highest wavenumber possible as the onset of a stop-band at $\beta_{sw} = \pi/d = 0.1k_0$ for the chosen unit cell discretization of $d = \lambda/20$ occurs at this wavenumber. This corresponds to a maximum sheet impedance of $-j20\Omega$ according to Fig. 4d. For this reason, hard limits of $-j20\Omega$ on the impedances during the optimization phase were set, and is why the tunnel sheet impedance is approximately $-j20\Omega$. The remaining evanescent spectrum is due to the sharp perturbations in the sheet impedances of Fig. 2c. These surface waves are responsible for redistributing power transversally within the port regions and at their transitions with the surface wave region in order to achieve passivity and losslessness.

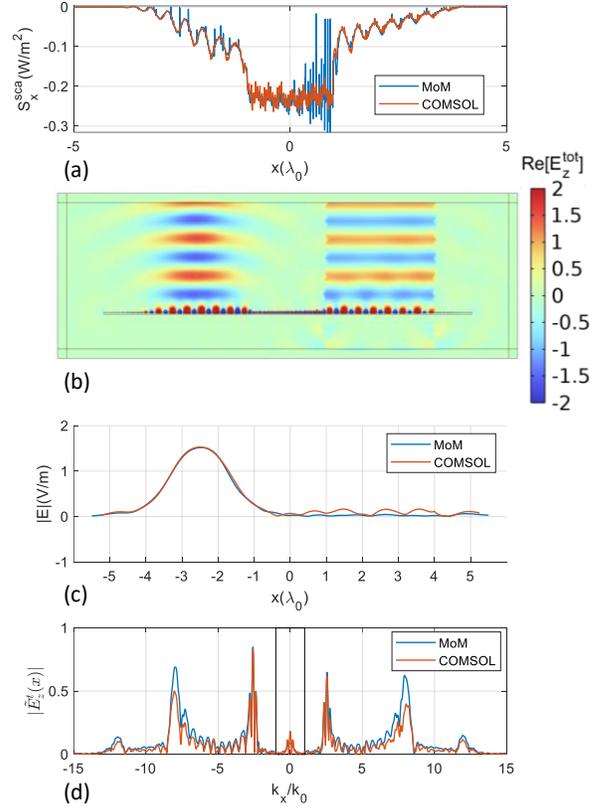

Fig. 6. COMSOL Multiphysics simulation results. (a) $x$-component of the Poynting vector at the plane of the metasurface. (b) Real part of the total electric field. (c) Line cut of near electric scattered field amplitude at a height of one wavelength above the metasurface. (d) Amplitude spectrum of the scattered electric field at the metasurface plane.

The power in the surface wave can be seen to grow approximately linearly in agreement with the conclusions in [3] in Fig. 5a, although here the spectrum (Fig. 5a) contains many spatial harmonics rather than the single harmonic considered in [3] and the metasurface is strongly non-local. The figure shows the $x$-component of the Poynting vector, $S_x^{sca} = -(1/2)Re[E_z^{sca}H_y^{sca*}]$. $H_y^{sca}$ was obtained by taking the inverse Fourier transform of $\tilde{E}_z^{sca}k_x/\eta_0k_0$, where $\tilde{E}_z^{sca}$ is the electric field spectrum at the plane of the metasurface (the amplitude of $\tilde{E}_z^{sca}$ is shown in Fig. 4a). The power density in the surface wave is shown to increase from zero within the input port region approximately linearly as more of the power in the plane wave is absorbed, then become constant through the tunnel region as the power is carried to the output port region, and finally decay approximately linearly in the output port region to zero as the power is shed into the scattered beam. The oscillations in the power density profile occur due to the interference between the similarly polarized incident and surface wave fields [16].

Next, to show the metasurface perfectly converts the incident plane wave at port 1 to the complex-valued scattered field at port 2, the near electric field was calculated along a horizontal line one wavelength above the metasurface. In Fig. 5b, the stipulated scattered field amplitude (replicated from Fig. 2a), the directly calculated (from the induced surface currents) scattered near field amplitude, and the backprojected far fields are all shown compared. It is evident that the non-local metasurface perfectly creates the stipulated near field amplitude, and hence transfers all power in the incident plane wave to the output scattered field. Integrating the power contained in the near fields along a horizontal line one wavelength above the metasurface yields $P_{sca,stip} = 0.12\text{mW/m}$ and $P_{sca,nf} = 0.1198\text{mW/m}$ giving a port-to-port transfer efficiency of 99%.

Lastly, to provide an independent verification of the teleporting metasurface, the design was imported into COMSOL Multiphysics and a full-wave simulation performed. The results are compared to the MoM results in Fig. 6. As can be seen, the independent full-wave verification corroborates our results.

## III. Conformal Teleporting Metasurfaces

By incorporating conformal geometry modelling capabilities into the integral equation/moment method algorithm [16], teleporting metasurfaces connecting two distant non-colinear ports in space can be accomplished. These types of teleporting metasurfaces can be useful for channel optimization in urban environments where the window-pane sized metasurface conforms to the corner of a building for example (see Fig. 7).

### A. Conformal Metasurface for Communications Channel Optimization

In Fig. 8, the geometry of a conformal teleporting metasurface which routes the surface wave around a 90° bend is shown. The metasurface is parameterized by a superquadric function with $p = 10$,

$$x(u,v) = \frac{v}{\sqrt[p]{\left(\frac{\cos u}{a}\right)^p + \left(\frac{\sin u}{b}\right)^p}}\cos u$$
$$y(u,v) = \frac{v}{\sqrt[p]{\left(\frac{\cos u}{a}\right)^p + \left(\frac{\sin u}{b}\right)^p}}\sin u$$
$$\left.\begin{array}{c}\pi/2 \le u \le \pi \\ 1 - d/a \le v \le 1\end{array}\right\} \quad (2)$$

The parameterization is also shown graphically in Fig. 9. The parameters $a$ and $b$ control the aperture length along the $x$-axis and $y$-axis, respectively, and the parameter $d$ controls the substrate thickness. The parameter $p$ controls the metasurface shape and radius of curvature at the bend. For $p = 2$, for example, (2) defines a quadrant of a circular annulus in the $xy$-plane. As $p \to \infty$, the parameterization approaches a quadrant of a square ring with thickness $d$. When $v = 1$, the superquadric has the largest radius (the curve $g$ in Fig. 9). The impedance sheet will be placed along this arc. When $v = 1 - d/a$, the superquadric has the smallest radius (the curve $e$ in Fig. 9). This is where the perfectly conducting ground plane will be placed. $\forall v$ between these two values, the space between is filled (the dielectric material of the substrate will fill this area). For the parameters in (2), the greatest radius of curvature at the 90° bend point is $R = 0.587\lambda_0$ [16]. The metasurface has length $a = 5\lambda_0$ along the $x$-axis, $b = 5\lambda_0$ along the $y$-axis, and thickness $d = 1.27\text{mm}$ (50mil). It is constructed from the same three layer stack: a patterned metallic cladding represented as a spatially variant homogenized impedance sheet, a dielectric spacer, and a ground plane. The incident plane wave has its $\vec{k}$ vector oriented along the $-y$-axis and illuminates the portion of the metasurface between $-4\lambda_0 \le x \le -\lambda_0$. The plane wave will be absorbed at this space wave port and converted into a surface wave. The surface wave will travel around the bend delivering the power to port 2 defined along $\lambda_0 \leq y \leq 4\lambda_0$, where it will be formed into a shaped beam corresponding to an aperture field with uniform amplitude and phase.

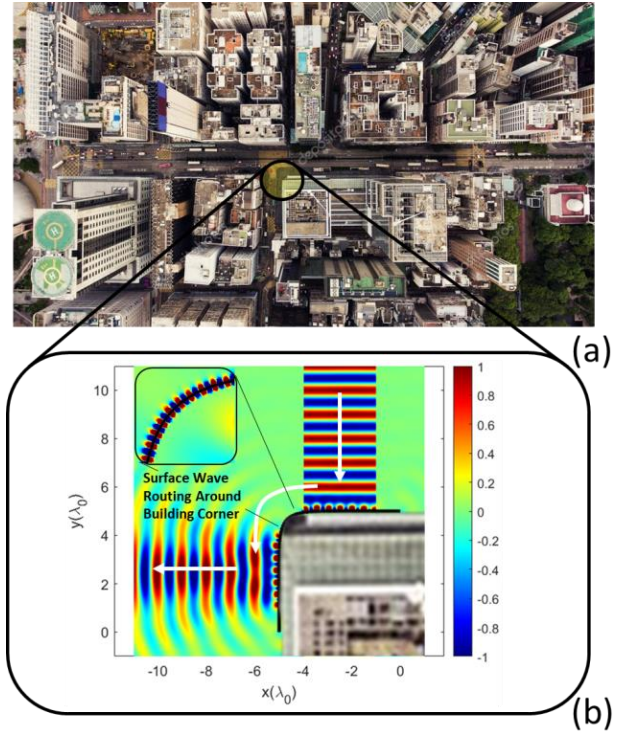

Fig. 7. (a) An urban environment. (b) Simulation results of a conformal teleporting metasurface which routes plane waves around corners of buildings.

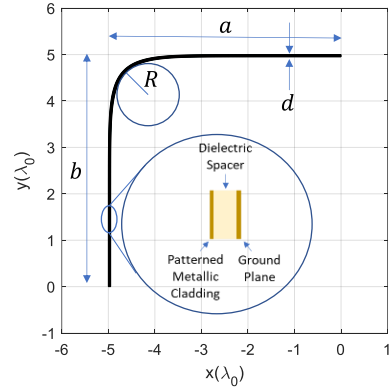

Fig. 8. Conformal teleporting metasurface geometry.

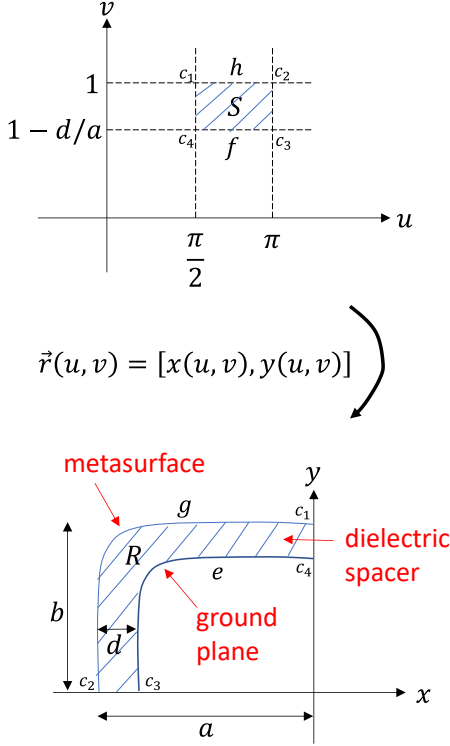

Fig. 9. Parameterization of conformal metasurface.

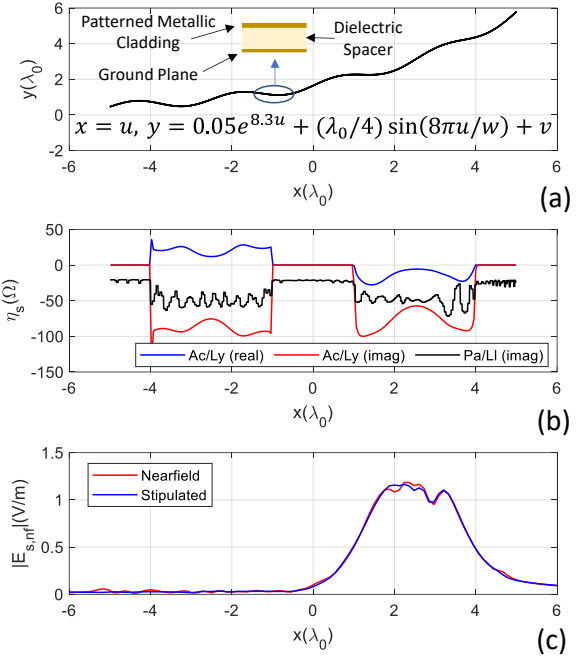

Fig. 11. (a) Conformal metasurface geometry. (b) Metasurface sheet impedances of initial local active/lossy (Ac/Ly) metasurface design, and subsequent non-local passive and lossless (Pa/Ll). (c) Line cut of near electric scattered field amplitude at a height of one wavelength above the metasurface.

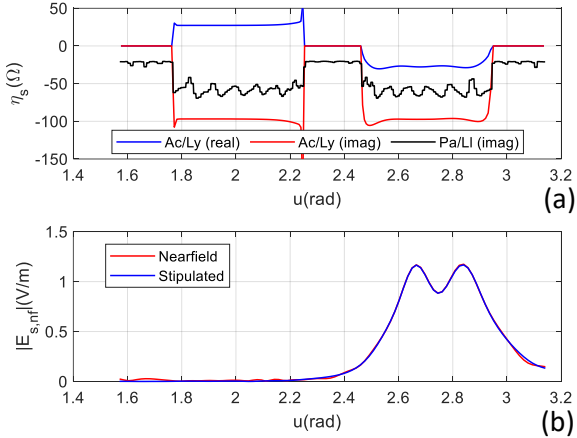

Fig. 10. (a) Metasurface sheet impedances of initial local active/lossy (Ac/Ly) metasurface design, and subsequent non-local passive/lossless (Pa/Ll) design vs. parameter $u$. (b) Line cut of near electric scattered field amplitude at a height of one wavelength above the metasurface.

Figure 10a shows the metasurface sheet impedances for both the local active/lossy metasurface design and the non-local passive/lossless metasurface design. Figure 10b shows the near field amplitude taken along a contour following the metasurface and one wavelength above the metasurface. As can be seen, the non-local passive/lossless metasurface performs identically to the local active/lossy design. Finally, the real part of the total near electric field is shown in Fig. 7b. As in the planar case, the metasurface is performing the function of perfect teleportation, only in this case, the beam is seen to teleport around the corner of a building.

### B. Sinusoidally Modulated Exponential Metasurface Coupler

The final example is a perfect conformal teleporting metasurface where the coupling regions are not planar. The geometry and its parameterization are shown in Fig. 11a and in Fig. 12, respectively [16]. The parameterization can be described as a sinusoidally modulated exponential

$$x(u,v) = u \qquad \qquad \qquad -w/2 \leq u \leq w/2$$
$$y(u,v) = ce^{au} + p\sin bu + v \qquad -d \leq v \leq 0 \qquad (3)$$

The parameter $w$ controls the aperture length along the $x$-axis, and the parameter $d$ controls the substrate thickness. The parameters $c$ and $a$ controls the amplitude and the growth rate of the exponential function, which acts as a fundamental term of which the sinusoid is added to. The parameters $p$ and $b$ control the amplitude and period of the sinusoidal term. The impedance sheet will be placed along the curve resulting from $v = 0$, The perfectly conducting ground plane will be placed along the curve at $v = -d$. $\forall v$ between these two values, the space between is filled (the dielectric material of the substrate will fill this area). The metasurface width, as projected onto the $x$-axis, is $w = 10\lambda_0$. The incident field is assumed a normally incident plane wave illuminating only the left-hand portion of the metasurface between $-4\lambda_0 \leq x \leq -1\lambda_0$. Defining the scattered field as $E_z^{sca} = e^{-jky}$, for $\lambda_0 \leq x \leq 4\lambda_0$, and solving the governing integral equation, results in the active/lossy design impedances shown in Fig. 11b. The corresponding passive/lossless design's reactances after optimization are also shown in Fig. 11b. Finally, the simulation results of the real part of the total near electric field for the excited passive/lossless design is shown in Fig. 13. Perfect teleportation is observed, as well as perfect coupling of a normally incident plane wave to a surface wave over a conformal surface.

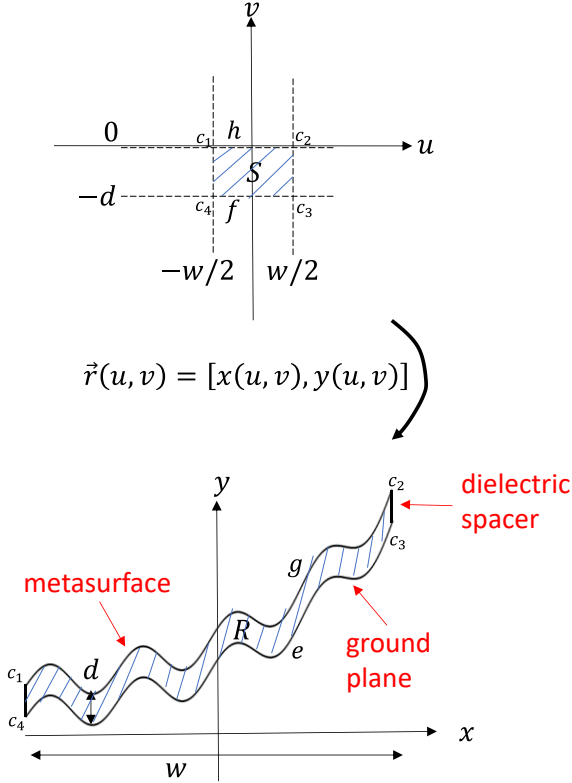

Fig. 12. Parameterization of conformal metasurface.

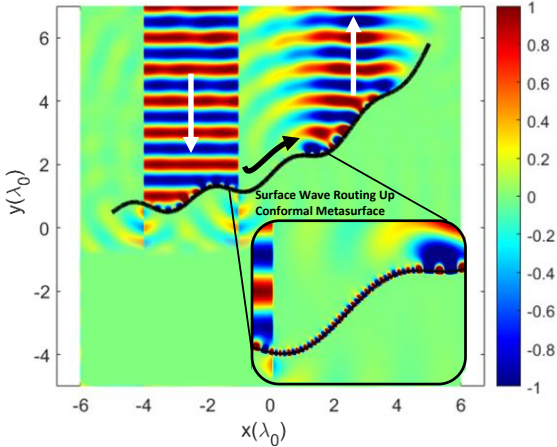

Fig. 13. Simulation results of a conformal teleporting metasurface shaped using a sinusoid added to an exponential function.

A COMSOL Multiphysics full-wave verification was also performed for this conformal metasurface design. The results are shown in Fig. 14. The figure shows some imperfect coupling and/or impedance matching between the port region and the tunnel region as some scattered electric field is present over the input port region. Nonetheless, the full-wave results again corroborate our results.

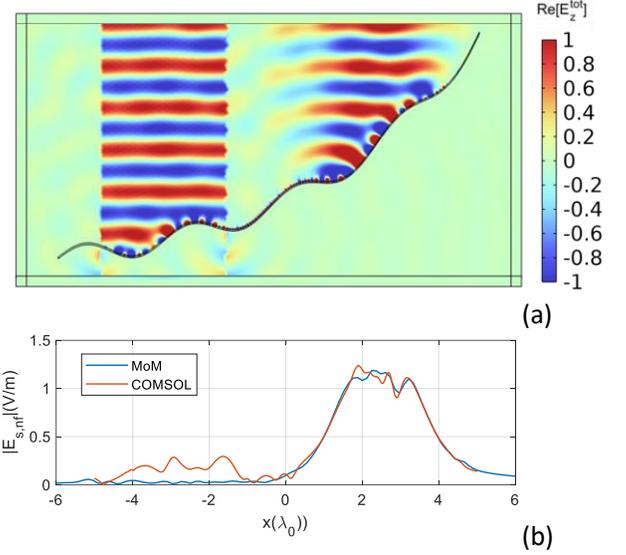

Fig. 14. COMSOL Multiphysics simulation results. (a) Real part of the total electric field. (c) Line cut of near electric scattered field amplitude at a height of one wavelength above the metasurface.

## V. CONCLUSION

Finally, note the primary purpose of this paper is to show that perfect teleportation is possible and what the sheet impedances look like. Support for dielectric materials is currently being added to the unit cell design process required to realize these metasurfaces outlined in [19]. Once this is complete, follow-on work involves translating the optimized impedance sheets for all designs in this paper to patterned metallic claddings.

## ACKNOWLEDGMENT

The author would like to acknowledge contributions by Professor Anthony Grbic and Dr. Luke Szymanski from the University of Michigan on previous related works.

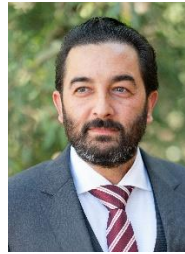

**Jordan Budhu** (Member, IEEE) Jordan Budhu received his M.S. degree in electrical engineering from the California State University, Northridge, California, USA, in 2010, and the Ph.D. degree in electrical engineering from the University of California, Los Angeles, California, USA, in 2018.

He is currently the Steven O. Lane Junior Faculty Fellow of Electrical and Computer Engineering in the Bradley Department of Electrical & Computer Engineering at Virginia Tech. Before being hired as Assistant Professor at Virginia Tech in 2022, he was a Postdoctoral Research Fellow in the Radiation Laboratory and a Lecturer in the Department of Electrical Engineering and Computer Science at the University of Michigan, Ann Arbor, Michigan, USA from 2019 to 2022. In 2011 and 2012, he was a Graduate Student Intern at the NASA Jet Propulsion Laboratory. In 2017, he was named a Teaching Fellow at the University of California, Los Angeles. His research interests are in metamaterials and metasurfaces, computational electromagnetics algorithms for metamaterial and metasurface design, conformal beamforming antennas, nanophotonics and metamaterials for the infrared, 3D printed inhomogeneous lens design, CubeSat antennas, reflectarray antennas, and antenna theory.

Dr. Budhu's awards and honors include the 2010 Eugene Cota Robles Fellowship from UCLA, the 2012 Best Poster award at the IEEE Coastal Los Angeles Class-Tech Annual Meeting, the 2018 UCLA Henry Samueli School of Engineering and Applied Science Excellence in Teaching Award, the first-place award for the 2019 USNC-URSI Ernst K. Smith Student Paper Competition at the 2019 Boulder National Radio Science Meeting, and the Steven O. Lane Junior Faculty Fellowship of Electrical and Computer Engineering in the Bradley Department of Electrical & Computer Engineering at Virginia Tech.

**Supplemental Material: Perfect Plane-Wave to Surface-Wave Coupler Enabled Teleporting Conformal Metasurfaces**

Jordan Budhu[1]

[1]Department of Electrical and Computer Engineering, Virginia Tech, Blacksburg, VA, USA

# 1 Conformal Method of Moments

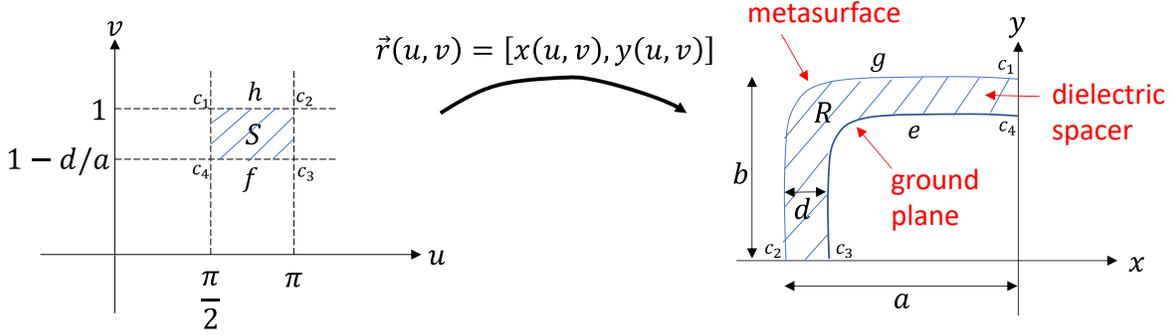

Fig. 1. Parameterization of conformal geometry.

## 1.1 Conformal Surface Parameterization

The conformal geometry is parameterized by the mapping function

$$\vec{r}:(u,v) \to (x,y) \qquad (1)$$

as shown in Fig. 1. Equation (1), in the present case, is defined by a super quadric function

$$
\left.
\begin{aligned}
x(u,v) &= \frac{v}{\sqrt[p]{\left(\dfrac{\cos u}{a}\right)^p + \left(\dfrac{\sin u}{b}\right)^p}} \cos u \\[2ex]
y(u,v) &= \frac{v}{\sqrt[p]{\left(\dfrac{\cos u}{a}\right)^p + \left(\dfrac{\sin u}{b}\right)^p}} \sin u
\end{aligned}
\right\}
\begin{aligned}
&\pi/2 \le u \le \pi \\
&1 - d/a \le v \le 1
\end{aligned}
\qquad (2)
$$

The parameters $a$ and $b$ control the aperture length along the $x$-axis and $y$-axis, respectively, and the parameter $d$ controls the substrate thickness. The parameter $p$ controls the metasurface shape and radius of curvature at the bend. For $p = 2$, for example, (2) defines a quadrant of a circular annulus in the $xy$-plane. As $p \to \infty$, the parameterization approaches a quadrant of a square ring with thickness $d$. When $v = 1$, the superquadric has the largest radius (the curve $g$ in Fig. 1). The impedance sheet will be placed along this arc. When $v = 1 - d/a$, the superquadric has the smallest radius (the curve $e$ in Fig. 1). This is where the perfectly conducting ground plane will be placed. $\forall v$ between these two values, the space between is filled (the dielectric material of the substrate will fill this area).

The derivatives of (2) are also useful. They are

$$\frac{\partial x}{\partial u} = \frac{-v}{\sqrt[p]{\left(\frac{\cos u}{a}\right)^p + \left(\frac{\sin u}{b}\right)^p}} \sin u$$

$$- \frac{v}{\left[\left(\frac{\cos u}{a}\right)^p + \left(\frac{\sin u}{b}\right)^p\right]^{\frac{1}{p}+1}} \left(\left(\frac{\cos u}{a}\right)^{p-1}\left(\frac{-\sin u}{a}\right) + \left(\frac{\sin u}{b}\right)^{p-1}\left(\frac{\cos u}{b}\right)\right)\cos u$$

$$\frac{\partial y}{\partial u} = \frac{v}{\sqrt[p]{\left(\frac{\cos u}{a}\right)^p + \left(\frac{\sin u}{b}\right)^p}} \cos u$$

$$- \frac{v}{\left[\left(\frac{\cos u}{a}\right)^p + \left(\frac{\sin u}{b}\right)^p\right]^{\frac{1}{p}+1}} \left(\left(\frac{\cos u}{a}\right)^{p-1}\left(\frac{-\sin u}{a}\right) + \left(\frac{\sin u}{b}\right)^{p-1}\left(\frac{\cos u}{b}\right)\right)\sin u$$

$$\frac{\partial x}{\partial v} = \frac{1}{\sqrt[p]{\left(\frac{\cos u}{a}\right)^p + \left(\frac{\sin u}{b}\right)^p}} \cos u$$

$$\frac{\partial y}{\partial v} = \frac{1}{\sqrt[p]{\left(\frac{\cos u}{a}\right)^p + \left(\frac{\sin u}{b}\right)^p}} \sin u$$

(3)

Also, the curvature, $\kappa$, of the impedance sheet layer can be computed from

$$\kappa(u) = \frac{\left|\vec{r}'(u,1) \times \vec{r}''(u,1)\right|}{\left|\vec{r}'(u,1)\right|^3}$$

$$\vec{r}' = \left[\frac{\partial}{\partial u}x(u,1), \frac{\partial}{\partial u}y(u,1)\right]$$

$$\vec{r}'' = \left[\frac{\partial^2}{\partial u^2}x(u,1), \frac{\partial^2}{\partial u^2}y(u,1)\right]$$

(4)

## 1.2 Integral Equations for Conformal Geometry

The conformal upgrade of the integral equation in [1]–[3] involves both surface and volume integrations over the boundary and volume of the domain $R$, respectively. Single integrals over curves bounding the region $R$ (curve $g$ or $e$ in Fig. 1 for example) can be evaluated by integrating over curves bounding the parameter space region $S$ (curve $h$ or $f$ in Fig. 1 for example) from

$$\int_{g,e} f(x,y)\,ds = \int_{h,f} f\big(x(u,v_0),y(u,v_0)\big)\sqrt{\left(\frac{\partial x}{\partial u}\Big|_{v=v_0}\right)^2+\left(\frac{\partial y}{\partial u}\Big|_{v=v_0}\right)^2}\,du \tag{5}$$

where $v_0 = 1$ or $1-d/a$ for integration over the metasurface layer (curves $g$ or $h$) or ground plane layer (curves $e$ or $f$), respectively. To evaluate the double integrals over the region $R$, the integrations can be done over the region $S$ in the parameter space using

$$\iint_R f(x,y)\,dA = \iint_S f\big(x(u,v),y(u,v)\big)\begin{vmatrix}\partial x/\partial u & \partial y/\partial u \\ \partial x/\partial v & \partial y/\partial v\end{vmatrix}du\,dv \tag{6}$$

Care must be taken to ensure the absolute value of the Jacobian determinant appearing in (6) is always taken (note, the Jacobian determinant associated with the mapping in Fig. 1 turns out to be negative so the absolute value must be taken).

The metasurface can be modeled as consisting of three layers. Layer 1, at $v_0 = 1$ denotes the impedance sheet. Layer 2, at $v_0 = v$, denotes the dielectric spacer. Layer 3, at $v_0 = 1-d/a$, denotes the perfectly conducting ground plane. An integral equation can be constructed for each layer, and hence there are three total integral equations, one for each choice of $v_0$ in the following

$$E^i\big(x(u,v_0),y(u,v_0)\big) = \eta_s\big(x(u,v_0),y(u,v_0)\big)J\big(x(u,v_0),y(u,v_0)\big)$$

$$+\frac{\eta_0 k_0}{4}\int_{u'=\pi/2}^{u'=\pi} H_0^{(2)}\left(k_0\sqrt{\big[x(u,v_0)-x(u',1)\big]^2+\big[y(u,v_0)-y(u',1)\big]^2}\right)$$

$$J\big(x(u',1),y(u',1)\big)\sqrt{\left(\frac{\partial x}{\partial u'}\Big|_{v'=1}\right)^2+\left(\frac{\partial y}{\partial u'}\Big|_{v'=1}\right)^2}\,du'$$

$$+\frac{\eta_0 k_0}{4}\int_{v'=1-d/a}^{v'=1}\int_{u'=\pi/2}^{u'=\pi} H_0^{(2)}\left(k_0\sqrt{\big[x(u,v_0)-x(u',v')\big]^2+\big[y(u,v_0)-y(u',v')\big]^2}\right)$$

$$J\big(x(u',v'),y(u',v')\big)\begin{vmatrix}\partial x/\partial u' & \partial y/\partial u' \\ \partial x/\partial v' & \partial y/\partial v'\end{vmatrix}du'\,dv'$$

$$+\frac{\eta_0 k_0}{4}\int_{u'=\pi/2}^{u'=\pi} H_0^{(2)}\left(k_0\sqrt{\big[x(u,v_0)-x(u',1-d/a)\big]^2+\big[y(u,v_0)-y(u',1-d/a)\big]^2}\right)$$

$$J\big(x(u',1-d/a),y(u',1-d/a)\big)\sqrt{\left(\frac{\partial x}{\partial u'}\Big|_{v'=1-d/a}\right)^2+\left(\frac{\partial y}{\partial u'}\Big|_{v'=1-d/a}\right)^2}\,du' \tag{7}$$

The three integral equations in (7) (one for each choice of $v_0$) can be simultaneously solved by the method of moments as presented in the next section.

## 1.3    Method of Moment Solution of Integral Equations

The current densities in (7) are expanded into pulse basis functions placed in the parametric space tessellating the region and boundary of $S$ (1D pulses for the boundary of $S$ mapping electric surface current densities on the metasurface and ground plane, and 2D pulses for the area within $S$ mapping polarization current densities in the dielectric substrate)

$$J_1(u) = J_1\big(x(u,1), y(u,1)\big) = \sum_{n=1}^{N_1} \alpha_n P_n\big(x(u,1), y(u,1)\big) \ , \ |u - u_n| \le \frac{\Delta u}{2}$$

$$J_2(u,v) = J_2\big(x(u,v), y(u,v)\big) = \sum_{n=1}^{N_2} \alpha_n P_n\big(x(u,v), y(u,v)\big) \ , \ |u - u_n| \le \frac{\Delta u}{2} \ , \ |v - v_n| \le \frac{\Delta v}{2}$$

$$J_3(u) = J_3\big(x(u,1-d/a), y(u,1-d/a)\big) = \sum_{n=1}^{N_3} \alpha_n P_n\big(x(u,1-d/a), y(u,1-d/a)\big) \ , \ |u - u_n| \le \frac{\Delta u}{2}$$

$$(8)$$

$\Delta u$ denotes the basis function width and $\Delta v$ their height in the 2D case. Substitution of (8) into (7) and testing the integral equation using the same expansion functions (Galerkin method) results in (for $v_0 = 1$)

$$\int_{u=u_m - \Delta u/2}^{u=u_m + \Delta u/2} E^i\big(x(u,1), y(u,1)\big) \sqrt{\left(\frac{\partial x}{\partial u}\Big|_{v=1}\right)^2 + \left(\frac{\partial y}{\partial u}\Big|_{v=1}\right)^2}\, du$$

$$= \eta_s\big(x(u,1), y(u,1)\big) \sum_{n=1}^{N_1} \alpha_n \int_{u=u_m - \Delta u/2}^{u=u_m + \Delta u/2} \delta_{mn} \sqrt{\left(\frac{\partial x}{\partial u}\Big|_{v=1}\right)^2 + \left(\frac{\partial y}{\partial u}\Big|_{v=1}\right)^2}\, du$$

$$+ \frac{\eta_0 k_0}{4} \sum_{n=1}^{N_1} \alpha_n \int_{u=u_m - \Delta u/2}^{u=u_m + \Delta u/2} \int_{u'=u_n - \Delta u/2}^{u'=u_n + \Delta u/2} H_0^{(2)}\left(k_0 \sqrt{[x(u,1) - x(u',1)]^2 + [y(u,1) - y(u',1)]^2}\right)$$

$$\sqrt{\left(\frac{\partial x}{\partial u'}\Big|_{v=1}\right)^2 + \left(\frac{\partial y}{\partial u'}\Big|_{v=1}\right)^2} \sqrt{\left(\frac{\partial x}{\partial u}\Big|_{v=1}\right)^2 + \left(\frac{\partial y}{\partial u}\Big|_{v=1}\right)^2}\, du'\, du$$

$$+ \frac{\eta_0 k_0}{4} \sum_{n=1}^{N_2} \alpha_n \int_{u=u_m - \Delta u/2}^{u=u_m + \Delta u/2} \int_{v'=v_n - \Delta v/2}^{v'=v_n + \Delta v/2} \int_{u'=u_n - \Delta u/2}^{u'=u_n + \Delta u/2} H_0^{(2)}\left(k_0 \sqrt{[x(u,1) - x(u',v')]^2 + [y(u,1) - y(u',v')]^2}\right)$$

$$\begin{vmatrix} \partial x/\partial u' & \partial y/\partial u' \\ \partial x/\partial v' & \partial y/\partial v' \end{vmatrix} \sqrt{\left(\frac{\partial x}{\partial u}\Big|_{v=1}\right)^2 + \left(\frac{\partial y}{\partial u}\Big|_{v=1}\right)^2}\, du'\, dv'\, du$$

$$+ \frac{\eta_0 k_0}{4} \sum_{n=1}^{N_3} \alpha_n \int_{u=u_m - \Delta u/2}^{u=u_m + \Delta u/2} \int_{u'=u_n - \Delta u/2}^{u'=u_n + \Delta u/2} H_0^{(2)}\left(k_0 \sqrt{[x(u,1) - x(u',1-d/a)]^2 + [y(u,1) - y(u',1-d/a)]^2}\right)$$

$$\sqrt{\left(\frac{\partial x}{\partial u'}\Big|_{v=1-d/a}\right)^2 + \left(\frac{\partial y}{\partial u'}\Big|_{v=1-d/a}\right)^2} \sqrt{\left(\frac{\partial x}{\partial u}\Big|_{v=1}\right)^2 + \left(\frac{\partial y}{\partial u}\Big|_{v=1}\right)^2}\, du'\, du$$

$$(9)$$

The integral equation in (9) can be written in the following form $V_1 = Z_{s1}I_1 + Z_{11}I_1 + Z_{12}I_2 + Z_{13}I_3 = (Z_{11} + Z_{s1})I_1 + Z_{12}I_2 + Z_{13}I_3$, where

$$[V_1]_{N_1 \times 1} = \int\limits_{u=u_m-\Delta u/2}^{u=u_m+\Delta u/2} E^i\left(x(u,1),y(u,1)\right) \sqrt{\left(\frac{\partial x}{\partial u}\bigg|_{v=1}\right)^2 + \left(\frac{\partial y}{\partial u}\bigg|_{v=1}\right)^2}\, du$$

$$[Z_{s1}]_{N_1 \times N_1} = \eta_s\left(x(u,1),y(u,1)\right) \int\limits_{u=u_m-\Delta u/2}^{u=u_m+\Delta u/2} \delta_{mn} \sqrt{\left(\frac{\partial x}{\partial u}\bigg|_{v=1}\right)^2 + \left(\frac{\partial y}{\partial u}\bigg|_{v=1}\right)^2}\, du$$

$$[Z_{11}]_{N_1 \times N_1} = \frac{\eta_0 k_0}{4} \int\limits_{u=u_m-\Delta u/2}^{u=u_m+\Delta u/2} \int\limits_{u'=u_n-\Delta u/2}^{u'=u_n+\Delta u/2} H_0^{(2)}\left(k_0\sqrt{\left[x(u,1)-x(u',1)\right]^2 + \left[y(u,1)-y(u',1)\right]^2}\right)$$
$$\sqrt{\left(\frac{\partial x}{\partial u'}\bigg|_{v=1}\right)^2 + \left(\frac{\partial y}{\partial u'}\bigg|_{v=1}\right)^2} \sqrt{\left(\frac{\partial x}{\partial u}\bigg|_{v=1}\right)^2 + \left(\frac{\partial y}{\partial u}\bigg|_{v=1}\right)^2}\, du'\,du$$

$$[Z_{12}]_{N_1 \times N_2} = \frac{\eta_0 k_0}{4} \int\limits_{u=u_m-\Delta u/2}^{u=u_m+\Delta u/2} \int\limits_{v'=v_n-\Delta v/2}^{v'=v_n+\Delta v/2} \int\limits_{u'=u_n-\Delta u/2}^{u'=u_n+\Delta u/2} H_0^{(2)}\left(k_0\sqrt{\left[x(u,1)-x(u',v')\right]^2 + \left[y(u,1)-y(u',v')\right]^2}\right)$$
$$\begin{vmatrix} \partial x/\partial u' & \partial y/\partial u' \\ \partial x/\partial v' & \partial y/\partial v' \end{vmatrix} \sqrt{\left(\frac{\partial x}{\partial u}\bigg|_{v=1}\right)^2 + \left(\frac{\partial y}{\partial u}\bigg|_{v=1}\right)^2}\, du'\,dv'\,du$$

$$[Z_{13}]_{N_1 \times N_3} = \frac{\eta_0 k_0}{4} \int\limits_{u=u_m-\Delta u/2}^{u=u_m+\Delta u/2} \int\limits_{u'=u_n-\Delta u/2}^{u'=u_n+\Delta u/2} H_0^{(2)}\left(k_0\sqrt{\left[x(u,1)-x(u',1-d/a)\right]^2 + \left[y(u,1)-y(u',1-d/a)\right]^2}\right)$$
$$\sqrt{\left(\frac{\partial x}{\partial u'}\bigg|_{v=1-d/a}\right)^2 + \left(\frac{\partial y}{\partial u'}\bigg|_{v=1-d/a}\right)^2} \sqrt{\left(\frac{\partial x}{\partial u}\bigg|_{v=1}\right)^2 + \left(\frac{\partial y}{\partial u}\bigg|_{v=1}\right)^2}\, du'\,du$$

$$[I_1]_{N_1 \times 1} = \alpha_{n1}$$
$$[I_2]_{N_2 \times 1} = \alpha_{n2}$$
$$[I_3]_{N_3 \times 1} = \alpha_{n3}$$

$$(10)$$

The self terms ($m = n$) in (10) are calculated using the procedure outlined in [4]. For completeness, the formulas for calculating the self-terms are provided in section 1.4. Following the same procedure for the remaining layers ($v_0 = v$ for layer 2 and $v_0 = 1 - d/a$ for layer 3) leads to the block matrix equation

$$\begin{bmatrix} V_1 \\ V_2 \\ V_3 \end{bmatrix} = \begin{bmatrix} Z_{11}+Z_{s1} & Z_{12} & Z_{13} \\ Z_{21} & Z_{22}+Z_v & Z_{23} \\ Z_{31} & Z_{32} & Z_{33} \end{bmatrix} \begin{bmatrix} I_1 \\ I_2 \\ I_3 \end{bmatrix} \qquad (11)$$

Note, for the dielectric layer, $Z_v = eye([j\omega\epsilon_0(\epsilon_r-1)]^{-1})$, where eye( ) indicates an identity matrix with the argument appearing along the diagonal. Also note, $\eta_s = 0$ for layer 3 and hence $Z_{s3}$ does not appear in (11). Finally, note in our implementation, each of the voltage vector and impedance matrix elements in (10) are normalized to their own arc length (for 1D) or area (for 2D).

The matrix equation (11) can be solved by either knowing the sheet impedances $Z_{s1}$ or by knowing the desired total field and making the substitution $W_1 = Z_{s1}I_1$ following from the boundary condition $E^{tot} = \eta_s J_1$

$$\begin{bmatrix} V - W_1 \\ V_2 \\ V_3 \end{bmatrix} = \begin{bmatrix} Z_{11} & Z_{12} & Z_{13} \\ Z_{21} & Z_{22} + Z_v & Z_{23} \\ Z_{31} & Z_{32} & Z_{33} \end{bmatrix} \begin{bmatrix} I_1 \\ I_2 \\ I_3 \end{bmatrix} \tag{12}$$

Since, in our case, the desired total field, $E^t$, is known, the $W_1$ vector can be found from

$$W_1 = \int\limits_{u=u_m-\Delta u/2}^{u=u_m+\Delta u/2} E^t\big(x(u,1), y(u,1)\big) \sqrt{\left(\frac{\partial x}{\partial u}\bigg|_{v=1}\right)^2 + \left(\frac{\partial y}{\partial u}\bigg|_{v=1}\right)^2}\, du \tag{13}$$

In this case, after solving equation (12) for the current $I_1$, the sheet impedances can be found by returning back to the boundary condition from $\eta_s = W_1/I_1$.

## 1.4    Calculation of Singular Matrix Element Terms

The self terms ($m = n$) in (10) are calculated using the procedure outlined in [4]. We summarize their results here adapted to our case of conformal method of moments. For more details, refer to the original paper [4]. There are two types of singular integrals, the surface integrals (associated with $[Z_{11}]$ and $[Z_{33}]$) and the volume integrals (associated with $[Z_{22}]$). We treat the surface integrals first.

### *Singular Surface Integrals*

The integral in question (for $[Z_{11}]$ for example) is

$$I_s = \frac{\eta_0 k_0}{4} \int\limits_{u=u_m-\Delta u/2}^{u=u_m+\Delta u/2} \int\limits_{u'=u_n-\Delta u/2}^{u'=u_n+\Delta u/2} H_0^{(2)}\left(k_0 \sqrt{[x(u,1)-x(u',1)]^2 + [y(u,1)-y(u',1)]^2}\right)$$

$$\sqrt{\left(\frac{\partial x}{\partial u'}\bigg|_{v=1}\right)^2 + \left(\frac{\partial y}{\partial u'}\bigg|_{v=1}\right)^2} \sqrt{\left(\frac{\partial x}{\partial u}\bigg|_{v=1}\right)^2 + \left(\frac{\partial y}{\partial u}\bigg|_{v=1}\right)^2}\, du'\, du \tag{14}$$

By singularity subtraction, the Hankel function can be written as

$$H_0^{(2)}\big(k_0 P\big) = \left[ H_0^{(2)}\big(k_0 P\big) + j\frac{2}{\pi}\ln P \right] - j\frac{2}{\pi}\ln P \tag{15}$$

where $P = \sqrt{\big(x(u,1)-x(u',1)\big)^2 + \big(y(u,1)-y(u',1)\big)^2}$. The first term in the brackets is well-behaved and can be integrated numerically. The strategy for the remaining term is to perform the inner integration analytically assuming a variable observation point $u$ passed in from the outer integral, then integrate the outer integral numerically using the analytic result from the inner integral as the integrand.

$$I_s = \frac{\eta_0 k_0}{4} \int\limits_{u=u_m-\Delta u/2}^{u=u_m+\Delta u/2} \int\limits_{u'=u_n-\Delta u/2}^{u'=u_n+\Delta u/2} \left[ H_0^{(2)}\big(k_0 P\big) + j\frac{2}{\pi}\ln P \right] \sqrt{\left(\frac{\partial x}{\partial u'}\bigg|_{v=1}\right)^2 + \left(\frac{\partial y}{\partial u'}\bigg|_{v=1}\right)^2} \sqrt{\left(\frac{\partial x}{\partial u}\bigg|_{v=1}\right)^2 + \left(\frac{\partial y}{\partial u}\bigg|_{v=1}\right)^2}\, du'\, du$$

$$- j\frac{\eta_0 k_0}{2\pi} \int\limits_{u=u_m-\Delta u/2}^{u=u_m+\Delta u/2} \left[ \int\limits_{u'=u_n-\Delta u/2}^{u'=u_n+\Delta u/2} \ln P \sqrt{\left(\frac{\partial x}{\partial u'}\bigg|_{v=1}\right)^2 + \left(\frac{\partial y}{\partial u'}\bigg|_{v=1}\right)^2}\, du' \right] \sqrt{\left(\frac{\partial x}{\partial u}\bigg|_{v=1}\right)^2 + \left(\frac{\partial y}{\partial u}\bigg|_{v=1}\right)^2}\, du$$

$$= I_{s1} + I_{s2}$$

$$\tag{16}$$

The inner integral of $I_{s2}$ (in brackets) to integrate analytically becomes

$$I_{s2,in}(u) = \int_{u'=u_n-\Delta u/2}^{u'=u_n+\Delta u/2} \ln P \sqrt{\left(\frac{\partial x}{\partial u'}\bigg|_{v=1}\right)^2 + \left(\frac{\partial y}{\partial u'}\bigg|_{v=1}\right)^2}\, du' \tag{17}$$

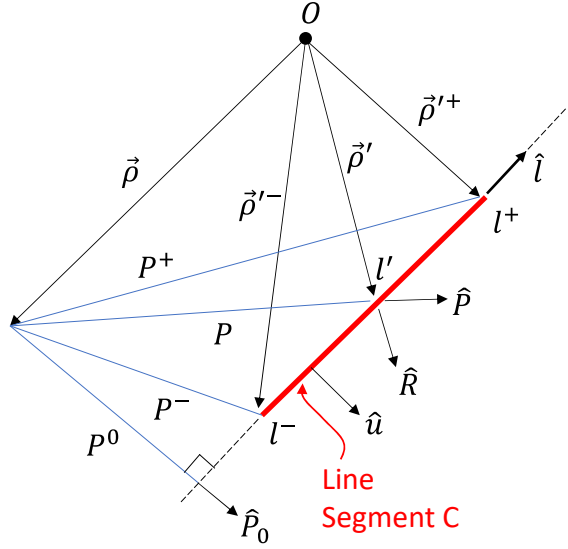

Fig. 2. Geometrical definitions for analytic line integration.

With reference to Fig. 2, the coordinate origin is denoted by $O$. The observation point position vector is denoted by $\vec{\rho} = [x(u,1), y(u,1)]$. The source point position vector is denoted by $\vec{\rho}' = [x(u',1), y(u',1)]$. The source line segment, $C$, is colored red in Fig. 2 and is parameterized by the arc length variable $l'$ measured from the line perpendicular to the extension of $C$ and which passes through the point located by $\vec{\rho}$. The endpoints of the source segment are pointed to by the vectors $\vec{\rho}'^-$ and $\vec{\rho}'^+$. The coordinates $P^0$ and $l'$ can be considered a pair of rectangular coordinates with origin at $\vec{\rho}$, locating points on the line segment $C$. Thus, the endpoints of the source segment are located at a distance of

$$P^\pm = \left|\vec{\rho}'^\pm - \vec{\rho}\right| = \sqrt{\left(P^0\right)^2 + \left(l^\pm\right)^2} \tag{18}$$

Other quantities in Fig. 2 are given by

$$\hat{l} = \frac{\vec{\rho}'^+ - \vec{\rho}'^-}{\left|\vec{\rho}'^+ - \vec{\rho}'^-\right|}$$

$$\hat{u} = \hat{l} \times \hat{n} = \hat{l} \times \hat{z}$$

$$l^\pm = \left(\vec{\rho}'^\pm - \vec{\rho}\right) \cdot \hat{l} \tag{19}$$

$$P^0 = \left|\left(\vec{\rho}'^\pm - \vec{\rho}\right) \cdot \hat{u}\right|$$

$$\hat{P}^0 = \frac{\left(\vec{\rho}'^\pm - \vec{\rho}\right) - l^\pm \hat{l}}{P^0}$$

All quantities in (18) and (19) can be solved for once $\vec{\rho}$ (the observation point) and $\vec{\rho}^\pm$ (the source segment endpoints) are defined. By defining

$$\vec{\rho} = [x(u,1), y(u,1)]$$
$$\vec{\rho}^{-} = [x(u_n - \Delta u/2, 1), y(u_n - \Delta u/2, 1)] \qquad (20)$$
$$\vec{\rho}^{+} = [x(u_n + \Delta u/2, 1), y(u_n + \Delta u/2, 1)]$$

the integration is done in the $xy$-space using the above paradigm rather than over the parametric $uv$-space since the vectors in (20) are constant (see Fig. 1). Thus, (17) becomes

$$I_{s2,in}(u) = \int_C \ln P\, dl' = \int_{l^-}^{l^+} \ln \sqrt{(P^0)^2 + (l')^2}\, dl'$$
$$= l^+ \ln P^+ - l^- \ln P^- + P^0 \left( \tan^{-1}\left(\frac{l^+}{P^0}\right) - \tan^{-1}\left(\frac{l^-}{P^0}\right) \right) - (l^+ - l^-)$$

$$(21)$$

The result in (21) can be singular if $P^{\pm} = 0$, which happens when the observation point $\vec{\rho}$ lies on either of the endpoints of the line segment $C$. In this case, the observation point can be set to $\vec{\rho} = \vec{\rho} + \epsilon \hat{u}$, where $\epsilon$ is a small constant. The complete integral (16) can now be evaluated numerically as

$$I_s = \frac{\eta_0 k_0}{4} \int_{u=u_m - \Delta u/2}^{u=u_m + \Delta u/2} \int_{u'=u_n - \Delta u/2}^{u'=u_n + \Delta u/2} \left[ 1 - j\frac{2}{\pi}\ln\left(\frac{\gamma}{2}k_0\right) \right] \sqrt{\left(\frac{\partial x}{\partial u'}\right)^2 + \left(\frac{\partial y}{\partial u'}\right)^2} \sqrt{\left(\frac{\partial x}{\partial u}\Big|_{v=1}\right)^2 + \left(\frac{\partial y}{\partial u}\Big|_{v=1}\right)^2} \, du' du$$

$$- j\frac{\eta_0 k_0}{2\pi} \int_{u=u_m - \Delta u/2}^{u=u_m + \Delta u/2} I_{s2,in}(u) \sqrt{\left(\frac{\partial x}{\partial u}\Big|_{v=1}\right)^2 + \left(\frac{\partial y}{\partial u}\Big|_{v=1}\right)^2} \, du$$

$$(22)$$

where $\gamma = 1.781$ and since $\left[ H_0^{(2)}(k_0 P) + j\frac{2}{\pi}\ln P \right] = 1 - j\frac{2}{\pi}\ln\left(\frac{\gamma}{2}k_0\right)$ by the small argument expansion for the Hankel function. A similar approach is used for the singular terms of $[Z_{33}]$. This completes the singular surface integral calculation. The singular volume integrals are handled next.

### Singular Volume Integrals

The integral in question is

$$I_v = \frac{\eta_0 k_0}{4} \int_{v=v_m - \Delta v/2}^{v=v_m + \Delta v/2} \int_{u=u_m - \Delta u/2}^{u=u_m + \Delta u/2} \int_{v'=v_n - \Delta v/2}^{v'=v_n + \Delta v/2} \int_{u'=u_n - \Delta u/2}^{u'=u_n + \Delta u/2} H_0^{(2)}\left( k_0 \sqrt{[x(u,v) - x(u',v')]^2 + [y(u,v) - y(u',v')]^2} \right)$$
$$\begin{vmatrix} \partial x/\partial u' & \partial y/\partial u' \\ \partial x/\partial v' & \partial y/\partial v' \end{vmatrix} \begin{vmatrix} \partial x/\partial u & \partial y/\partial u \\ \partial x/\partial v & \partial y/\partial v \end{vmatrix} du' dv' du dv$$

$$(23)$$

Using the singularity subtraction technique, (23) becomes

$$I_v = \frac{\eta_0 k_0}{4} \int_{v=v_m - \Delta v/2}^{v=v_m + \Delta v/2} \int_{u=u_m - \Delta u/2}^{u=u_m + \Delta u/2} \int_{v'=v_n - \Delta v/2}^{v'=v_n + \Delta v/2} \int_{u'=u_n - \Delta u/2}^{u'=u_n + \Delta u/2} \left[ 1 - j\frac{2}{\pi}\ln\left(\frac{\gamma}{2}k_0\right) \right] \begin{vmatrix} \partial x/\partial u' & \partial y/\partial u' \\ \partial x/\partial v' & \partial y/\partial v' \end{vmatrix} \begin{vmatrix} \partial x/\partial u & \partial y/\partial u \\ \partial x/\partial v & \partial y/\partial v \end{vmatrix} du' dv' du dv$$

$$- j\frac{\eta_0 k_0}{2\pi} \int_{v=v_m - \Delta v/2}^{v=v_m + \Delta v/2} \int_{u=u_m - \Delta u/2}^{u=u_m + \Delta u/2} \int_{v'=v_n - \Delta v/2}^{v'=v_n + \Delta v/2} \int_{u'=u_n - \Delta u/2}^{u'=u_n + \Delta u/2} \ln P \begin{vmatrix} \partial x/\partial u' & \partial y/\partial u' \\ \partial x/\partial v' & \partial y/\partial v' \end{vmatrix} \begin{vmatrix} \partial x/\partial u & \partial y/\partial u \\ \partial x/\partial v & \partial y/\partial v \end{vmatrix} du' dv' du dv$$

$$= I_{v1} + I_{v2}$$

$$(24)$$

where $P = \sqrt{\left(x(u,v) - x(u',v')\right)^2 + \left(y(u,v) - y(u',v')\right)^2}$. Integral $I_{v1}$ can be integrated numerically. Integral $I_{v2}$ will be integrated analytically for the inner integral and numerically for the outer integral.

$$I_{v2} = -j\frac{\eta_0 k_0}{2\pi} \int\limits_{v=v_m-\Delta v/2}^{v=v_m+\Delta v/2} \int\limits_{u=u_m-\Delta u/2}^{u=u_m+\Delta u/2} \left[ \int\limits_{v'=v_n-\Delta v/2}^{v'=v_n+\Delta v/2} \int\limits_{u'=u_n-\Delta u/2}^{u'=u_n+\Delta u/2} \ln P \begin{vmatrix} \partial x/\partial u' & \partial y/\partial u' \\ \partial x/\partial v' & \partial y/\partial v' \end{vmatrix} du'dv' \right] \begin{vmatrix} \partial x/\partial u & \partial y/\partial u \\ \partial x/\partial v & \partial y/\partial v \end{vmatrix} dudv$$

$$(25)$$

Again, the term in brackets will be labeled the inner integral, $I_{v2,in}$. To integrate $I_{v2,in}$ analytically, we employ a Gauss integral theorem. Thus, in the $xy$-space, we express the integrand as

$$I_{2v,in} = \int\limits_A \ln P dx'dy' = \lim_{\varepsilon \to 0}\frac{1}{2}\int\limits_{A-A_\varepsilon} \nabla'_s \bullet \left[ \left( P\ln P - \frac{P}{2} \right)\hat{P} \right] dx'dy' + \lim_{\varepsilon \to 0}\int\limits_{A_\varepsilon} \ln P dx'dy' \quad (26)$$

where $A_\epsilon$ is a small circular region of radius $\epsilon$ enclosing the observation point included to make the integrand continuously differentiable. For more information, see [4]. A divergence theorem can be applied to (26) to express the surface integral in terms of a flux integral around the boundary enclosing the area $A$. Following the result derived in [4], (26) is evaluated as

$$I_{v2,in}(u,v) = \frac{1}{2}\sum_{i=1}^{4}\vec{P}_i^0 \bullet \hat{u}_i \left[ l_i^+ \ln P_i^+ - l_i^- \ln P_i^- + P_i^0\left( \tan^{-1}\left(\frac{l_i^+}{P_i^0}\right) - \tan^{-1}\left(\frac{l_i^-}{P_i^0}\right) \right) - \frac{3}{2}\left( l_i^+ - l_i^- \right) \right] \quad (27)$$

where $\vec{P}_i^0 = P_i^0 \hat{P}_i^0$, $\hat{P} = (\vec{\rho}' - \vec{\rho})/P$, and $\hat{u}$ is the outward normal vector (ensure $\hat{z}\cdot(\hat{u}_i \times \hat{l}_i) > 0$, else set $\hat{u}_i = -\hat{u}_i$). Note, the mapping in (1) negates the signed area (Jacobian determinant is negative), and thus, going clockwise around the perimeter of the quadrilateral patch in the $uv$-space maps to a counterclockwise path around the perimeter of $A$ in the native $xy$-space (see Fig. 3). Index $i$ denotes one of the four sides of the quadrilateral patch in $R$ bounding the surface area $A$, and thus each side $i$ is represented by the geometry in Fig. 2.

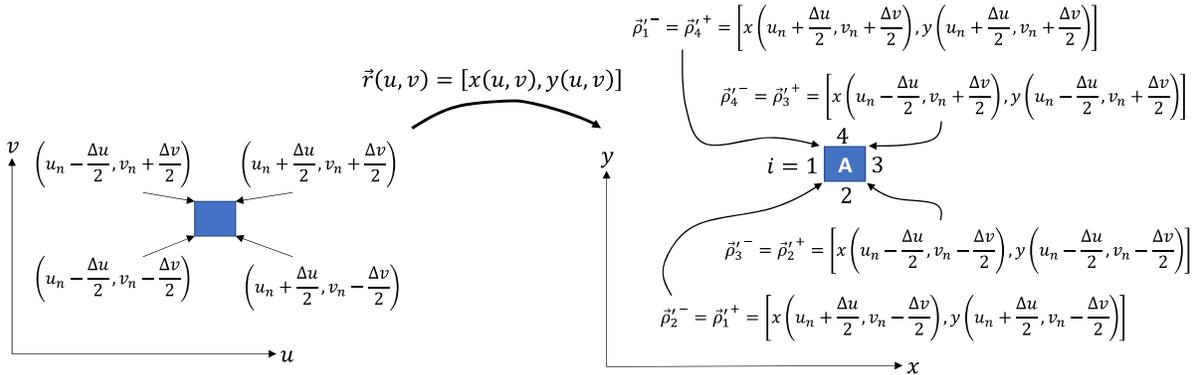

Fig. 3. Mapping of a quadrilateral patch in $uv$-space to its corresponding quadrilateral patch A in the $xy$-space.

Note, although the quadrilateral patch in the $uv$-space is rectangular (is constructed from straight line segments), its image under the mapping $\vec{r}$ may have curved bounding line segments. However, if the discretization is fine enough, the perimeter of $A$ can be approximated as constructed from straight line segments, and the results of Fig. 2 will hold. With this result, (24) is found as

$$I_v = \frac{\eta_0 k_0}{4} \int\limits_{v=v_m - \Delta v/2}^{v=v_m + \Delta v/2} \int\limits_{u=u_m - \Delta u/2}^{u=u_m + \Delta u/2} \int\limits_{v'=v_n - \Delta v/2}^{v'=v_n + \Delta v/2} \int\limits_{u'=u_n - \Delta u/2}^{u'=u_n + \Delta u/2} \left[1 - j\frac{2}{\pi}\ln\left(\frac{\gamma}{2}k_0\right)\right] \begin{vmatrix} \partial x/\partial u' & \partial y/\partial u' \\ \partial x/\partial v' & \partial y/\partial v' \end{vmatrix} \begin{vmatrix} \partial x/\partial u & \partial y/\partial u \\ \partial x/\partial v & \partial y/\partial v \end{vmatrix} du' dv' du dv$$

$$- j\frac{\eta_0 k_0}{2\pi} \int\limits_{v=v_m - \Delta v/2}^{v=v_m + \Delta v/2} \int\limits_{u=u_m - \Delta u/2}^{u=u_m + \Delta u/2} I_{v2,in}(u,v) \begin{vmatrix} \partial x/\partial u & \partial y/\partial u \\ \partial x/\partial v & \partial y/\partial v \end{vmatrix} du dv$$

(28)

This completes the singular term evaluations.

## 2 Parameterization for SinExp Teleporting Metasurface

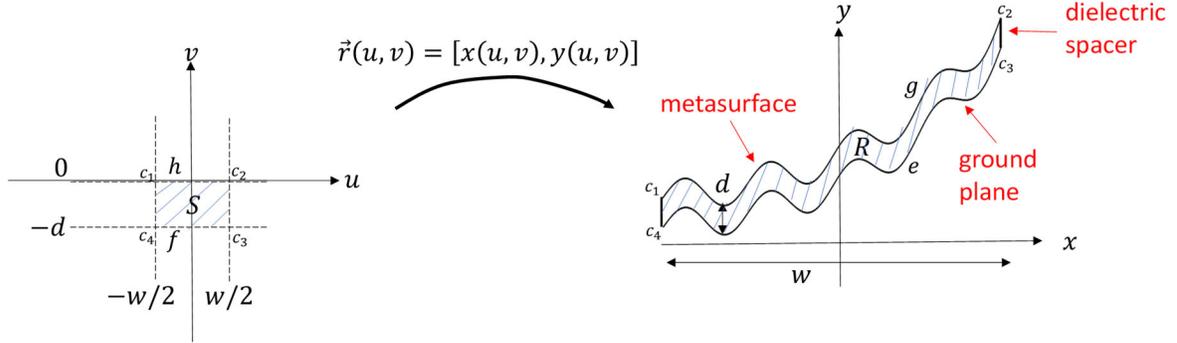

Fig. 4. Parameterization of conformal geometry.

### 2.1 Conformal Surface Parameterization

The conformal geometry is parameterized by the mapping function

$$\vec{r}:(u,v) \rightarrow (x,y)$$

(29)

as shown in Fig. 4. Equation (29), in this case, is defined by the function

$$\left.\begin{array}{l} x(u,v) = u \\ y(u,v) = ce^{au} + p\sin bu + v \end{array}\right\} \begin{array}{l} -w/2 \leq u \leq w/2 \\ -d \leq v \leq 0 \end{array}$$

(30)

The parameter $w$ controls the aperture length along the $x$-axis, and the parameter $d$ controls the substrate thickness. The parameters $c$ and $a$ controls the amplitude and the growth rate of the exponential function, which acts as a fundamental term of which the sinusoid is added. The parameters $p$ and $b$ control the amplitude and period of the sinusoidal term. The impedance sheet will be placed along the curve resulting from $v = 0$, The perfectly conducting ground plane will be placed along the curve at $v = -d$. $\forall v$ between these two values, the space between it is filled (the dielectric material of the substrate will fill this area). The derivatives of (2) are also useful. They are

$$\frac{\partial x}{\partial u} = 1, \qquad \frac{\partial y}{\partial u} = cae^{au} + pb\cos bu, \qquad \frac{\partial x}{\partial v} = 0, \qquad \frac{\partial y}{\partial v} = 1$$

(31)

The integral equations for this case are the same as in section 1.

## 3 Normally Incident Plane Wave Coupled Into a Single Harmonic Surface Wave

If the incident field is a normally incident plane wave

$$\vec{E}^i = \hat{z} E_0 e^{jk_0 y} \tag{32}$$

and the desired scattered field is a surface wave of wavenumber $\beta_s$ and attenuation constant $\alpha_x$ in the direction of surface wave propagation and $\alpha_y = \sqrt{\beta_s^2 - k_0^2}$ normal to the metasurface

$$\vec{E}^s = A e^{-(\alpha_x + j\beta_s)x - \alpha_y y} \tag{33}$$

it can be shown that the metasurface required to obtain the transformation must contain loss and gain or be non-local. To show this, we first find the associated magnetic fields

$$H_y^i = \frac{1}{j\omega\mu}\frac{\partial E_z^i}{\partial x} = 0$$

$$H_x^i = \frac{-1}{j\omega\mu}\frac{\partial E_z^i}{\partial y} = -\frac{E_z^i}{\eta_0}$$

$$H_y^s = \frac{1}{j\omega\mu}\frac{\partial E_z^s}{\partial x} = \frac{-(\alpha_x + j\beta_s)}{j\omega\mu}E_z^s \tag{34}$$

$$H_x^s = \frac{-1}{j\omega\mu}\frac{\partial E_z^s}{\partial y} = \frac{\alpha_y}{j\omega\mu}E_z^s$$

The Poynting vector can be calculated at the metasurface plane

$$\vec{S} = \frac{1}{2}\text{Re}\left[\vec{E}\times\vec{H}^*\right] = \frac{1}{2}\text{Re}\left[\left(E_z^i\hat{z} + E_z^s\hat{z}\right)\times\left(H_x^i\hat{x} + H_x^s\hat{x} + H_y^s\hat{y}\right)^*\right]$$

$$= \frac{1}{2}\text{Re}\left[\hat{y}\left(E_z^i H_x^{i*} + E_z^i H_x^{s*} + E_z^s H_x^{i*} + E_z^s H_x^{s*}\right) + \hat{x}\left(-E_z^s H_y^{s*} - E_z^i H_y^{s*}\right)\right] \tag{35}$$

Separating it into $\hat{x}$ and $\hat{y}$ components gives

$$S_y = -\frac{\left|E_z^i\right|^2}{2\eta_0} - \frac{1}{2\eta_0}\text{Re}\left[E_z^s E_z^{i*}\right]$$

$$S_x = \frac{\beta_s}{2\omega\mu}\left|E_z^s\right|^2 + \frac{1}{2}\text{Re}\left[\frac{\beta_s}{\omega\mu}E_z^i E_z^{s*} + j\frac{\alpha_x}{\omega\mu}E_z^i E_z^{s*}\right] \tag{36}$$

The interference terms (second term on the right hand side of both of (36)) lead to oscillations in the power density profile. For a local pointwise passive and lossless metasurface, $S_y = 0$, however, the first term on the right hand side of the first of (36) is constant whereas the second term on the right hand side is an oscillating function of $x$, hence $S_y$ cannot be made zero. To nullify $S_y$, the metasurface will either need local loss and gain to compensate the oscillations and null the normal power density or contain surface waves to shuttle power from places where local loss is needed to places where local gain is needed. In this case, the metasurface is considered non-local. The second of (36) also shows the transverse power density is an oscillating function of $x$ as in Fig. 5a of the paper.